# Exceptional points and spectral singularities in active epsilon-near-zero plasmonic waveguides

Ying Li and Christos Argyropoulos*

Dept. of Electrical & Computer Engineering, University of Nebraska-Lincoln, Lincoln, NE, 68588, USA

*christos.argyropoulos@unl.edu

*The intriguing physics of exceptional points and spectral singularities in open (non-Hermitian) active photonic systems has recently sparked increased interest among the research community. These spectral degeneracies have been obtained in asymmetric active and passive photonic configurations but their demonstration with symmetric active plasmonic structures still remains elusive. In this work, we present a nanoscale active plasmonic waveguide system consisting of an array of periodic slits that can exhibit exceptional points and spectral singularities leading to several novel functionalities. The proposed symmetric active system operates near its cut-off wavelength and behaves as an effective epsilon-near-zero (ENZ) medium. We demonstrate the formation of an exceptional point (EP) that is accessed with very low gain coefficient values, a unique feature of the proposed nanoscale symmetric plasmonic configuration. Reflectionless ENZ transmission and perfect loss-compensation are realized at the EP which coincides with the effective ENZ resonance wavelength of the proposed array of active plasmonic waveguides. When we further increase the gain coefficient of the dielectric material loaded in the slits, a spectral singularity occurs at the ENZ resonance leading to super scattering (lasing) response at both forward and backward directions. These interesting effects are achieved by materials*

*characterized by very small gain coefficients with practical values and at subwavelength scales due to the strong and homogeneous field enhancement inside the active slits at the ENZ resonance leading to enhanced light-matter interaction. We theoretically analyze the obtained EP, as well as the divergent spectral singularity, using a transmission-line model and investigate the addition of a second incident wave and nonlinearities in the response of the proposed active ENZ plasmonic system. Our findings provide a novel route towards interesting nanophotonic applications, such as reflectionless active ENZ media, unidirectional coherent perfect absorbers, nanolasers, and strong optical bistability and all-optical switching nanodevices.*

**1. Introduction**

Exceptional points (EPs) are branch point singularities of the spectrum found in open non-Hermitian Hamiltonian systems that correspond to the coalescence of both eigenvalues and eigenvectors [1]. They have been observed in parity-time (PT) symmetric non-Hermitian photonic systems and have led to several counterintuitive phenomena including unidirectional light propagation [2–5], loss-induced lasing [6], topological energy transfer [7], and asymmetric mode switching [8]. Not limited to PT-symmetric structures, EPs have also been detected in other open non-Hermitian systems without gain, such as metallic gratings [9], photonic crystal slabs [10], asymmetric plasmonic nanostructures [11], lossy metamaterials [12], and graphene [13]. All of these interesting systems exhibit some form of geometrical or material asymmetry in order to break parity reversal symmetry and achieve the EP response. A unique characteristic of EPs is that the reflection of the system becomes zero from different directions, which provides a strong indication of potential EP formation in the pursuit of their detection. Besides EPs, another class of irregular spectral points obtained only in active systems is the so-called spectral singularities that correspond to a lack of completeness of the system's eigenfunctions in a continuous spectrum [14]. They lead to distinct spectral degeneracies where the reflection and transmission of the system tend to infinity and have extremely narrow resonant bandwidth [15]. This interesting feature can be used for potential applications relevant to super scattering [16,17], coherent perfect absorption and lasing [18–20], and saturable nonlinearity [21]. Note that the majority of the previous works are focused on asymmetric nanoscale structures and the realization of EPs in symmetric plasmonic configurations have not been presented so far.

In parallel with the aforementioned research efforts, realistic metamaterials with effective epsilon-near-zero (ENZ) permittivity have attracted considerable attention in recent years due to their extraordinary optical transmission response characterized by infinite phase velocity combined with large and uniform field confinement [22–29]. Some interesting works have recently been proposed [30–34] that connected active ENZ metamaterials, even zero index metamaterials (ZIM), with PT-symmetry in order to realize the formation of EPs and spectral singularities or other degeneracies. However, the permittivity of the loss material in these systems needs to be equal to the conjugate of the permittivity of the gain medium to obtain efficient PT-symmetric effects, such as the EP formation. Note that, in the case of plasmonic nanoscale systems, it is very difficult to find optical active (gain) media with a positive imaginary part of permittivity equal to the usually large, especially for metals, optical losses associated to the negative imaginary part of the loss material permittivity [35]. Fortunately, the response of plasmonic waveguides operating near their cut-off frequency exhibiting effective ENZ response is accompanied by a uniform and strong field enhancement inside their nanochannels, where the gain material is located. This unique property is expected to trigger EPs and other PT-symmetric effects with minimum gain values, since the enhanced electromagnetic mode is evenly spread throughout the entire gain dielectric medium leading to a strong interaction with it. Hence, arrays of ENZ waveguides are anticipated to bring the formation of EPs and spectral singularities in symmetric plasmonic nanoscale systems one step closer to their practical implementation.

In this work, we propose a realistic design of a non-Hermitian nanophotonic system based on an array of ENZ waveguides loaded with active dielectric materials with low gain values with the goal to obtain PT-symmetric effects with symmetric nanostructures. In the past, it was demonstrated that this plasmonic configuration exhibits an effective ENZ response at its cut-off frequency and Fabry-Pérot (FP) resonances at higher frequencies [23–25,36]. A more detailed analysis about the ENZ response of the proposed system at its linear operation can be found in [36]. In this work, we study the effect of low gain in this system, which is induced by realistic active materials loaded inside the symmetric plasmonic nanoslits. Our aim is to completely compensate the loss of the proposed plasmonic ENZ configuration and achieve more complicated optical effects derived by PT-symmetry. Towards these goals, we numerically demonstrate and analytically prove the existence of an EP spectral degeneracy at the ENZ resonance that leads to reflectionless transparency combined with total loss compensation at the ENZ frequency point. In addition, a spectral singularity with super scattering (lasing) response is obtained by the same compact ENZ nanoscale system when active materials with slightly larger gain coefficients are introduced inside the nanochannels. The proposed active optical ENZ system has several additional unique features. For example, it is demonstrated that it can be used to dynamically modulate the output ENZ power from phase-controlled amplification to unidirectional absorption when the waveguide is illuminated by two counter-propagating plane waves. Finally, the proposed active ENZ configuration is also found to drastically boost several relative weak optical nonlinear effects, such as optical bistability and all-optical switching [37,38]. It is interesting that all these effects are achieved by a purely symmetric plasmonic configuration that does not break parity reversal symmetry. Our findings provide a new platform to design lossless ENZ metamaterials, low threshold nanolasers, and unidirectional coherent perfect absorbers [39].

## 2. Numerical Modeling

The geometry of the proposed plasmonic non-Hermitian nanophotonic system is illustrated in Fig. 1. It is composed of an array of narrow periodic rectangular slits loaded with an active dielectric material carved in a silver screen. The loss is introduced in the proposed non-Hermitian system by the silver screen, in addition to the inherent radiation losses of the system. The silver permittivity dispersion values are taken by previously derived experimental data [40]. The slits are loaded with an active dielectric material exhibiting gain with a relative permittivity at steady-state equal to $\varepsilon = \varepsilon_r + i\delta$, where the real part is equal to $\varepsilon_r = 2.2$ and the imaginary part corresponds to the gain coefficient that, for now, is set to be an arbitrary variable with constant value $\delta$. This is a valid approximation to model active materials with low gain coefficient values at the steady-state condition [41,42]. In addition, always an $\exp(i2\pi ft)$ time convention is assumed in the current work. The slit dimensions are chosen to have width $w = 200\,\text{nm}$, height $t = 40\,\text{nm}$ ($t \ll w$), and thickness $l = 500\,\text{nm}$, respectively, and the grating period is selected to be equal to $a = b = 400\,\text{nm}$. This free-standing waveguide geometry can sustain an ENZ resonance at the cut-off wavelength of its dominant quasi-TE$_{10}$ mode. [25,36]. It has been proven that extraordinary optical transmission can occur at the ENZ wavelength combined with large field enhancement and uniform phase distribution inside the nanoslits [28,36]. This is due to an anomalous matching phenomenon that depends only on the interface properties, i.e., on the aperture to period ratio of the array, and is therefore independent to the grating periodicity and each waveguide channel thickness.

The active plasmonic waveguide system is illuminated by a *z*-polarized plane wave, as schematically illustrated in Fig. 1. Note that all the simulations presented in this manuscript are performed by using the RF module of COMSOL Multiphysics [43], which is a numerical simulation software based on the Finite Element Method (FEM). It is used to solve the linear and nonlinear Maxwell's equations and model the electromagnetic response of the proposed active ENZ plasmonic waveguides. More details about the used modelling method can be found in [36]. The contour plots of transmittance and reflectance as a function of the incident wavelength and the small values of the active dielectric permittivity imaginary part $\delta$ are computed and shown in Figs. 2(a) and 2(b), respectively [36]. The wavelength in these plots varies very close to the ENZ cut-off frequency point of the system that was found to be approximately equal to $\lambda = 1011$nm in the case of passive slits ($\delta = 0$). Interestingly, reflectionless ENZ response is observed for a very low gain value of $\delta = 0.011$ at $\lambda = 1011$nm [point A in Fig. 2(b)]. This is clearly shown in Fig. 3, where the computed transmittance and reflectance of both passive ($\delta = 0$) and active ($\delta = 0.011$) systems are plotted. In the absence of gain ($\delta = 0$ / black lines), resonant optical transmission occurs at two different frequency points: the ENZ cut-off wavelength $(\lambda = 1011\text{nm})$ and the FP resonance $(\lambda = 922\,\text{nm})$. The transmittances have relatively small values (*T*~0.5) for these two wavelength points and cannot reach unity due to strong ohmic losses coming from the metallic (silver) walls of the waveguides. However, when an active dielectric material with small gain coefficient equal to $\delta = 0.011$ is included in the waveguide channels (red lines in Fig. 3), perfect transmittance and zero reflectance are obtained at the ENZ wavelength point. This response offers a clear indication of an EP formation which is now obtained by a purely symmetric active nanoplasmonic structure. The proposed system is completely different than the usually used asymmetric active photonic microscale structures exhibiting PT-symmetry [3].

The corresponding electric field enhancement distribution at the reflectionless ENZ wavelength point A is shown in Fig. 2(c) along a cross-section of the waveguide unit cell. The fields inside the waveguide, plotted in Fig. 2(c), are homogeneous and enhanced, a distinct characteristic of ENZ response. In this case, the transmitted field (right side of waveguide) has almost the same field values as the incident field (left side of waveguide), which means that there is no reflection and all the incident energy is perfectly transmitted through the matched nanochannel. We will prove in the next section 3 that, indeed, an EP is formed at this wavelength (point A) where perfect loss-compensated ENZ response is achieved which is not affected by the losses of the plasmonic waveguide. The loss and gain parameters of the resistive metallic waveguides and the active dielectric material are perfectly balanced at this ENZ EP. Note that in the proposed open non-Hermitian system the radiation loss also needs to be included in the total loss of the system, in addition to the loss stemming from the plasmonic material. The presented perfect loss-compensated ENZ response can be used to further improve the enhancement of quantum effects, such as superradiance [28], and boost optical nonlinearities [25] by skipping the usual loss limitations of plasmonic systems. The EP response at the ENZ wavelength is relative similar to the unidirectional reflectionless transparency obtained before for larger microscale (not nanoscale) PT-symmetric photonic systems [3,44]. However, in the currently proposed design the zero reflection response is obtained from both sides due to the symmetric structure of the proposed active plasmonic waveguide system. In addition and even more importantly, in the proposed unique ENZ configuration we obtain this effect by using a subwavelength active plasmonic configuration with an extremely thin thickness of just $l = 500\,\text{nm}$.

Next, we further increase the imaginary part of the gain material permittivity loaded in the nanochannels. When a gain value of $\delta = 0.038$ is reached, the response of the proposed open non-Hermitian plasmonic system is totally different compared to the lower gain EP response presented before and is transformed to a spectral singularity [16]. A super scattering (lasing) response is observed at the spectral singularity point B ($\delta = 0.038$, $\lambda = 1011.7\,\text{nm}$) in Fig. 2(b), obtained in the vicinity of the passive system's ENZ resonance. Giant and narrow transmittance and reflectance are simultaneously achieved by the proposed active plasmonic system at point B, which are ideal conditions for nanolasing and sensor applications. The corresponding fields inside the active waveguides also diverge and the surrounding fields are enhanced by many orders of magnitude. The calculated normalized electric field enhancement distribution in this case is depicted in Fig. 2(d). It is expected that the inherent nonlinear response of the gain material used in the proposed active system will eventually saturate the spectral singularity [21]. This study is outside of the scope of the current work and will be performed in the future.

The gain loaded inside the nanochannels has a constant non-dispersive value and our next step will be to consider a more practical gain material model. To this end, we include the realistic frequency dispersion in the gain material that follows Kramers-Kronig relations. Hence, the dielectric properties of the active medium are modeled with the Lorentz dispersion model with relative permittivity: $\varepsilon = \varepsilon_\infty + \varepsilon_{Lorentz}\omega_0^2 / \left(\omega_0^2 - 2i\omega\delta_0 - \omega^2\right)$, where $\varepsilon_\infty = 2.175$, $\varepsilon_{Lorentz} = 0.06325$,

$\omega_0 = 4.2 \times 10^{15}$ rad/s, $\delta_0 = 5.0 \times 10^{15}$ rad/s and $\omega = 2\pi f$ [39,45]. Realistic gain materials that exhibit these dispersion values are dyes (for example Rhodamine) doped in dielectric materials [46,47]. The corresponding active material disperson curves around the operating wavelength are plotted in Fig. 4(a). The real permittivity in this case is equal to $\varepsilon_r \approx 2.2$ and the imaginary part (gain) is approximately $\delta \approx 0.038$, which are ideal values to achieve the spectral singularity effect that was mentioned in the previous paragraph. The computed transmittance and reflectance of the active system loaded with the presented realistic gain material are plotted in Fig. 4(b), where the black lines refer to the plasmonic channels loaded with a passive dielectric material ($\delta = 0$, also shown in Fig. 3), while the red lines represent the channels loaded with the realistic active material characterized by the aforementioned Lorentzian dispersion model. As already mentioned, in the absence of the gain medium (black lines), resonant optical transmission occurs at the ENZ cut-off wavelength $(\lambda = 1011\,\text{nm})$ and the FP resonance $(\lambda = 922\,\text{nm})$. The maximum transmission is accompanied by minimum reflection at these wavelength points that consists a typical resonance behavior.

In the passive case, the summation of transmittance $T$ and reflectance $R$ is always less than one $(T + R < 1)$, since the presented array of passive waveguides is a lossy system and part of the incident energy is absorbed by the ohmic losses of the waveguides' metallic walls. However, this situation changes dramatically when the realistic active dielectric material with Lorentzian dispersion is included in the waveguide channel. An ultrasharp spectral singularity with giant and narrow transmittance and reflectance is obtained, interestingly, only at the ENZ resonance [red lines in Fig. 4(b)] and not at the FP resonance. This response is similar to the one obtained in point B of Fig. 2 for the constant (not dispersive) gain coefficient. Hence, super scattering and lasing with a very small practical gain coefficient is achieved only at the ENZ resonance frequency point. This is in stark contrast to the FP resonance response $(\lambda = 922\,\text{nm})$ of the same active system loaded with the same gain coefficient material, also shown by the red lines in Fig. 4(b), where there is no super scattering mainly due to the inhomogeneous and relative weak fields inside the plasmonic nanochannels.

## 3. Theoretical Analysis

The simulation results presented in the previous section provide a clear indication of EP formation and super scattering (lasing) response. The presented reflectionless transparency caused by the EP and the obtained divergent super scattering properties are further analyzed and scrutinized by using a transmission-line analytical model. This simple model can provide additional physical insights to the currently proposed active plasmonic system and can accurately verify the previously presented simulation results. The unit cell of the ENZ plasmonic waveguides section is described by a transmission line segment with length $l$, wavenumber $\beta_{wg}$, and characteristic impedance $Z_{wg}$. It is schematically depicted in Fig. 5(a) and is surrounded by free space with a constant impedance equal to $Z_0 = 377\,\Omega$. The guided wavenumber $\beta_{wg}$ of the ENZ mode, which supports a dominant quasi-TE$_{10}$ mode at the proposed plasmonic system, is calculated by solving the dispersion equation [26]:

$$\tan\left(\sqrt{\beta_{pp}^2 - \beta_{wg}^2}\,\frac{w}{2}\right) = \frac{\sqrt{\beta_{wg}^2 - k_{Ag}^2}}{\sqrt{\beta_{pp}^2 - \beta_{wg}^2}}, \tag{1}$$

where $k_{Ag} = k_0\sqrt{\varepsilon_{Ag}}$ is the wavenumber of silver with relative Drude permittivity dispersion $\varepsilon_{Ag} = \varepsilon_\infty - f_p^2/\left[f(f-i\gamma)\right]$, $f_p = 2175\text{THz}$, $\gamma = 4.35\text{THz}$, $\varepsilon_\infty = 5$ [40], $k_0$ is the free space wavenumber, and $\beta_{pp}$ is the guided wavenumber in an equivalent parallel-plate waveguide with the same height $t$ but infinite width $w$ that can be computed by the following dispersion formula [48]:

$$\tanh\left(\sqrt{\beta_{pp}^2 - \varepsilon k_0^2}\,\frac{t}{2}\right) = -\frac{\varepsilon}{\varepsilon_{Ag}}\frac{\sqrt{\beta_{pp}^2 - k_{Ag}^2}}{\sqrt{\beta_{pp}^2 - \varepsilon k_0^2}}, \qquad (2)$$

where $\varepsilon = \varepsilon_r + i\delta$ is the relative permittivity of the active dielectric material loaded inside the nanochannels. Thus the characteristic impedance of the dominant quasi-TE mode inside the plasmonic waveguide operating at the ENZ wavelength is defined as [49]:

$$Z_{wg} = \frac{\mu_0 \omega}{\beta_{wg}}, \qquad (3)$$

where $\omega = 2\pi f$ is the radial frequency. The effective permittivity of the active ENZ waveguide system can also be derived, in the limit $\varepsilon_{Ag} \to -\infty$ (assuming perfectly conducting metal) and $\beta_{pp} = k_0\sqrt{\varepsilon_r}$, to be equal to [25,36]: $\varepsilon_{eff} = \varepsilon_r - \frac{\pi^2}{k_0^2 w^2} + i\delta$. The real part of this formula will become equal to zero at the ENZ resonance for a waveguide width equal to $w = \frac{\pi}{\sqrt{\varepsilon_r}k_0}$, which is the classic cut-off condition of rectangular waveguides operating at microwave frequencies.

The proposed plasmonic configuration is a two-port network. In this case, we can use the guided wavenumber $\beta_{wg}$ and characteristic impedance $Z_{wg}$ to construct the $2\times 2$ ABCD matrix, which can relate the total electric and magnetic fields at the output and input ports of this system. This matrix is written as [50]:

$$\begin{bmatrix} A & B \\ C & D \end{bmatrix} = \begin{bmatrix} \cos(\beta_{wg}l) & iZ_{wg}\sin(\beta_{wg}l) \\ iZ_{wg}^{-1}\sin(\beta_{wg}l) & \cos(\beta_{wg}l) \end{bmatrix}. \quad (4)$$

By computing the ABCD parameters, we can calculate all the elements of the scattering matrix **S** that directly correspond to the reflection $(S_{11}, S_{22})$ and transmission $(S_{21}, S_{12})$ coefficients seen from both directions of the system [50]:

$$\begin{aligned} S_{11} &= \frac{A + B/Z_0 - CZ_0 - D}{\Delta} \\ S_{12} &= \frac{2(AD - BC)}{\Delta} \\ S_{21} &= \frac{2}{\Delta} \\ S_{22} &= \frac{-A + B/Z_0 - CZ_0 + D}{\Delta} \end{aligned}, \quad (5)$$

where $\Delta = A + B/Z_0 + CZ_0 + D$ and $Z_0 = 377\Omega$ is the surrounding free space impedance. The currently system is symmetric and reciprocal leading to $S_{11} = S_{22}$ and $S_{21} = S_{12}$.

Based on the aforementioned transmission-line method, we analytically compute the three-dimensional (3D) distribution of the reflectance $|S_{11}|^2$ versus the incident wavelength and gain coefficient $\delta$ in the vicinity of the ENZ resonance. The results are demonstrated in Fig. 5(b), where two interesting points with very high (divergent) and low (approximately zero) reflectance values are observed in the vicinity of the ENZ resonance depending on the value of the gain parameter $\delta$. The analytical results are consistent with the previously demonstrated simulation results shown in Fig. 2(b) and prove that the presented transmission line model can accurately predict the performance of the proposed active ENZ nanophotonic system.

To gain even more physical insights, we further analyze the impedance matching conditions where the reflectionless ENZ and super scattering frequency points occur. The free space surrounding the plasmonic waveguide is represented by two parallel lumped impedance elements $Z_0$ connected to the transmission line segment of the waveguide, as it can be seen in Fig. 5(a). The input impedance $Z_{in}$ seen looking towards the back load $Z_0$ of the plasmonic waveguide system at a distance $l$ from this load is given by [50]:

$$Z_{in} = Z_{wg} \frac{Z_0 + iZ_{wg}\tan(\beta_{wg}l)}{Z_{wg} + iZ_0\tan(\beta_{wg}l)}. \tag{6}$$

The reflection coefficient of the entire plasmonic ENZ system is analytically calculated by $r = (Z_{in} - Z_0)/(Z_{in} + Z_0)$, due to the front load $Z_0$. The transmission coefficient of the same configuration is equal to $t = 2Z_{in}/(Z_{in} + Z_0)$ [50]. In order to obtain perfect transmission and reflectionless responses ($r$=0 and $t$=1) at the ENZ EP, the real part [Re($Z_{in}$)] and imaginary part [Im($Z_{in}$)] of the system's input impedance have to satisfy the following perfect impedance matching relationships:

$$\begin{aligned}\operatorname{Re}(Z_{in}) &= Z_0 \\ \operatorname{Im}(Z_{in}) &= 0\end{aligned} \tag{7}$$

because $Z_0 = 377\Omega$ takes always real values. We plot in Fig. 5(c) the real and imaginary part of the input impedance $Z_{in}$ versus the incident wavelength for a fixed gain coefficient equal to $\delta = 0.017$. It is interesting that exactly at the ENZ resonance $(\lambda = 1016\,\text{nm})$, the input impedance $Z_{in}$ perfectly satisfies the conditions given by Eqs. (7). The presented impedance matching process results in perfect ENZ transmission combined with reflectionless ENZ propagation, as it was clearly demonstrated in Fig. 5(b).

Similarly, in order to achieve simultaneously diverging reflection and transmission coefficients (super scattering), the denominators of *r* and *t* need to become equal to zero. Therefore, the real and imaginary part of the system's input impedance should satisfy the following lasing conditions:

$$\begin{aligned}\text{Re}(Z_{in}) &= -Z_0 \\ \text{Im}(Z_{in}) &= 0\end{aligned}. \qquad (8)$$

These interesting conditions are also verified by plotting the real and imaginary part of the input impedance, now for a higher gain coefficient $\delta = 0.039$, with results depicted in Fig. 5(d). Note that the results obtained by the proposed analytical transmission line method are very close to the previously presented in section 2 simulation results. Some minor discrepancies are expected and are mainly due to the numerical approximations that are inevitable made during the simulations.

Next, we compute the eigenvalues and eigenvectors of the proposed plasmonic ENZ system to unambiguously prove that the presented reflectionless perfect ENZ transmission happens, indeed, at an EP. Towards this goal, we introduce the transfer matrix **M** formalism, which provides a direct relationship between the outgoing and input waves of the system. The eigenvalues of the transfer matrix **M** of the proposed plasmonic ENZ structure are computed by [16]:

$$\eta_{1,2} = \frac{M_{11} + M_{22}}{2} \pm \sqrt{\left(\frac{M_{11} + M_{22}}{2}\right)^2 - 1} \,, \qquad (9)$$

where the transfer matrix components $M_{11}$ and $M_{22}$ are calculated by using the transfer and scattering matrix elements relationship [51]:

$$\mathbf{M} = \begin{bmatrix} M_{11} & M_{12} \\ M_{21} & M_{22} \end{bmatrix} = \begin{bmatrix} S_{21} - \dfrac{S_{22}S_{11}}{S_{12}} & \dfrac{S_{22}}{S_{12}} \\ -\dfrac{S_{11}}{S_{12}} & \dfrac{1}{S_{12}} \end{bmatrix}. \qquad (10)$$

Equation (10) is derived if we assume that the determinant of the transfer matrix is unitary $\det(\mathbf{M}) = 1$ and the condition $M_{12} = -M_{21}$ is satisfied, which are valid relationships for reciprocal and symmetric systems, similar to the current case [16]. Hence, the product of the two eigenvalues of the system always needs to satisfy the relation $|\eta_1 \eta_2| = 1$, meaning that either each eigenvalue is unimodular or the eigenvalues form pairs with reciprocal moduli [19]. In contrast to the orthogonal eigenvectors of a Hermitian system, the eigenvectors of a non-Hermitian Hamiltonian system are biorthogonal given by the relation [16]:

$$\langle \varphi_1^l | \varphi_2^r \rangle = \langle \varphi_2^l | \varphi_1^r \rangle = 0, \qquad (11)$$

where the left $\varphi_{1,2}^l$ and right $\varphi_{1,2}^r$ eigenvectors of the transfer matrix $\mathbf{M}$ and are computed by the following formulas:

$$\langle \varphi_{1,2}^l | = \left( \dfrac{\eta_{1,2} - M_{22}}{M_{12}} \quad 1 \right)$$

$$|\varphi_{1,2}^r\rangle = \begin{pmatrix} \dfrac{\eta_{1,2} - M_{22}}{M_{21}} \\ 1 \end{pmatrix} . \qquad (12)$$

We present in Figs. 6(a) and 6(b) the evolution of the eigenvalues computed by Eq. (9) in the complex domain as a function of the gain coefficient $\delta$ at the ENZ resonance $(\lambda = 1016\,\text{nm})$ and slightly off the ENZ wavelength $(\lambda = 1015.8\,\text{nm})$, respectively. Interestingly, a clear EP degeneracy is observed for $\delta=0.017$ at the ENZ cut-off resonance wavelength $(\lambda = 1016\,\text{nm})$ [Fig. 5(a)], which is manifested by the collapse of the two eigenvalues into one along the unit circle only for this particular gain coefficient value. On the contrary, the two eigenvalues are always separated and there is no EP formation at a wavelength slightly off the ENZ resonance [Fig. 6(b)].

We also compute and present in Fig. 6(c) the absolute values of the two eigenvalues, calculated again by using Eq. (9), as a function of the incident wavelength for two different gain coefficients: $\delta=0.017$ (solid lines) and $\delta=0.04$ (dashed lines). In the case of $\delta=0.017$, a clear bifurcation point is observed at the ENZ resonance $(\lambda = 1016\,\text{nm})$. The observed bifurcation point again coincides with the ENZ wavelength which is another clear indication of an EP formation. When the wavelength is smaller than the ENZ resonance, the computed eigenvalues are unimodular, implying that there is no net amplification nor dissipation. However, when the wavelength is larger than the resonance, the eigenvalues have reciprocal moduli, with one magnitude greater than one and the other less than one, corresponding to states of amplification and dissipation, respectively [16]. In between and exactly at the ENZ resonance wavelength, there is a clear bifurcation point which is equivalent to an EP formed only on the ENZ response of this particular ENZ system. On the contrary, in the case of higher gain $\delta=0.04$, the computed eigenvalues are not unimodular and the EP degeneracy does not exist anymore. In addition, we have verified that when encircling the EP in the complex frequency space, the eigenvalues will be interchanged after a complete loop, due to the square root singularity behavior of the EP [52–55]. Finally, the poles of the presented system have been checked in the complex frequency space and lead to a stable EP response.

The associated eigenvectors, computed by using Eq. (12), also coalesce at the ENZ resonance for a very small gain coefficient of $\delta=0.017$, which is another unambiguous indication of an EP, and the results are shown in Fig. 6(d). Therefore, the ohmic losses induced by the metallic waveguides in addition to the radiation losses of the proposed open non-Hermitian Hamiltonian system are completely compensated by the active dielectric material leading to the formation of an EP degeneracy, which, interestingly, is obtained by a purely symmetric nanoscale plasmonic configuration. The formation of the corresponding EP results in a totally lossless (or loss-compensated) ENZ response leading to the reflectionless transparency behavior shown in Fig. 5(b). Both eigenvalues and associated eigenvectors coalesce at the presented bifurcation point that is analytically proven to be certainly an EP with intriguing properties [16].

In addition, the other interesting spectral degeneracy point in Fig. 5(b) exhibits divergent reflectance values and satisfies the condition $M_{22}=0$ at the ENZ wavelength for slightly larger gain values $\delta=0.039$. This is a necessary condition in order to obtain a spectral singularity leading to the calculation of a gain threshold value that will cause lasing operation [15,20]. However, we have computed that $M_{11}\neq 0$ at the same ENZ wavelength, indicating that the anti-lasing (also known as coherent perfect absorption) condition cannot be satisfied at the same frequency point if we excite the plasmonic structure with two counter propagating beams [20]. Therefore, as the gain coefficient $\delta$ is increased from 0.017 to 0.039, the proposed open non-Hermitian active plasmonic system is transformed into a system where gain dominates its response; the transitional point (EP) is transformed to a spectral singularity, which is manifested as a super scattering (or lasing) response of the system. Interestingly, the transmission and reflection of the system decrease again to low values in the case of further gain increase above the lasing threshold (not shown here), a different response compared to typical lasing system configurations.

## 4. Applications

In this section, we use the proposed active ENZ plasmonic system to investigate some interesting applications stemming from its unique properties, including unidirectional absorption and the enhancement of third-order optical nonlinear effects. The same active plasmonic waveguides system, which exhibits super scattering spectral degeneracy and is characterized by the previously presented Lorentzian model [Fig. 4(a)], is illuminated by two counter-propagating plane waves with equal intensities $I_0$ and arbitrary phase difference $\Delta\psi=\psi_2-\psi_1$, as illustrated in Fig. 7(a). The computed phase dependent output power in port 1 (Output 1) and port 2 (Output 2) as a function of the phase difference $\Delta\psi$ is demonstrated in Fig. 7(b) when the incident wavelengths of the two beams are fixed slightly off the ENZ resonance $(\lambda=1009\,\text{nm})$. Interestingly, almost perfect directional absorption is obtained only from one side of the waveguide (either Output 1 or 2), surprisingly by a perfectly symmetric structure, when a phase offset $\Delta\psi$ equal to either 148° or 212° is used.

The output power from both sides is modulated from amplification (super scattering) to perfect directional absorption just by carefully tuning the phase offsets between the two counter-propagating waves. It is noteworthy that the strong modulation between high to almost zero transmission happens with a purely reciprocal symmetric system with nanoscale subwavelength dimensions. The computed normalized electric field distribution at one of the directional absorption phase difference conditions $\left(\Delta\psi = 212°\right)$ is shown in the inset of Fig. 7(b), where it can be seen that the power is mainly scattered unidirectional to the output port 1 and the transmission is almost zero at output port 2. Note that the field distribution inside the nanochannel is homogeneous and strong because we operate very close to the ENZ resonance. Such an efficient control of the radiation direction, achieved in subwavelength scale, is unprecedented. It is envisioned to have several all-optical switching and routing applications in integrated nanophotonic components and nanocircuits. Finally, we would like to stress that the currently proposed plasmonic system cannot work as a coherent perfect absorber under the illumination of two counter-propagating waves and can only work as a laser [15,20]. The complex transmission and reflection coefficients *t* and *r* do not have equal amplitudes $\left(|t| \neq |r|\right)$, even at the ENZ super scattering point shown in Fig. 4(b), and the phase difference between transmission and reflection coefficients cannot be made equal to either 0 or $\pm\pi$ [51,56]. These critical conditions are absolutely needed in order to achieve coherent perfect absorption and cannot be realized with the currently proposed active ENZ plasmonic system.

The strong and uniform fields inside the waveguide channels at the ENZ EP [shown in Fig. 2(c)] are expected to naturally cause the proposed plasmonic system to access the optical nonlinear regime as the intensity values of the input illumination are increased. The triggered boosted third-order nonlinearity will make the system become intensity-dependent and bistable. Due to this reason, we study the third-order Kerr optical nonlinear effect in the active plasmonic waveguide system illuminated now by only one plane wave incident from one side [36]. In this case, the slits in Fig. 1 are loaded with a Kerr nonlinear material with relative permittivity $\varepsilon_{ch} = \varepsilon + \chi^{(3)} |\mathbf{E}_{ch}|^2$, where $\varepsilon = 2.2 + 0.011i$ represents the active dielectric material with a gain coefficient $\delta = 0.011$ corresponding to the response demonstrated by point A in Fig. 2(b), $\chi^{(3)} = 4.4 \times 10^{-20} \, \text{m}^2/\text{V}^2$ is a typical third-order nonlinear dielectric material susceptibility [57], and $\mathbf{E}_{ch}$ is the local electric field induced inside the nanochannels.

We illuminate the waveguides with a *z*-polarized plane wave (similar to Fig. 1) and compute the transmittance versus the wavelength for a fixed input intensity $I_0 = 1500 \, \text{MW}/\text{cm}^2$. The result is shown in Fig. 8(a), where we also present in the same figure the linear operation (black line) to better illustrate the effect of nonlinearity to the proposed ENZ system. In the case of linear operation, perfect reflectionless transmission is obtained at the ENZ EP ($\lambda = 1011 \, \text{nm}$), similar to point A in Fig. 2(b). In the nonlinear case (red lines), a strong optical bistablity or broad hysteresis appears which exhibits two stable transmission branches. The peak of the upper transmission branch reaches very close to one values (total transmission) and, at the same time, the corresponding lower transmission branch reaches zero values. Perfect transmission (ON state) or total reflection (zero transmission-OFF state) is achieved with the same nonlinear structure just by increasing or decreasing the incident radiation's wavelength (shown by the arrows in Fig. 8). This interesting property consists the epitome of all-optical nonlinear switching behavior. Figure 8(b) demonstrates the variation in the transmission versus the input intensity $I_0$ for two different incident radiation wavelengths, both chosen to be slightly larger than the ENZ EP wavelength. Two broad nonlinear transmission hysteresis loops exist under relative low threshold input intensities (approximately $1000\,\text{MW}/\text{cm}^2$) and the peak transmission in both cases is almost one (total transmission). Hence, the proposed optical nonlinear active ENZ plasmonic design is an ideal platform to be used for tunable all-optical switching applications [25,58,59]. Note that the combination of nonlinearities and non-Hermitian Hamiltonian systems has been demonstrated before mainly with microscale bulk photonic structures, especially for nonlinear PT-symmetric photonic systems [60], but these interesting effects have not been presented yet with subwavelength nonlinear plasmonic systems.

Finally, it is interesting to mention that the proposed active ENZ plasmonic system is able to slow down the incident light at the ENZ EP, where the group velocity nearly vanishes [61]. The group velocity is defined as $v_g = \left(d\beta_{wg}/d\omega\right)^{-1}$, where the guided wavenumber $\beta_{wg}$ at the ENZ wavelength of the proposed array of plasmonic waveguides is calculated by using Eq. (1). In the case of usual lossy ENZ plasmonic waveguides ($\delta = 0$), slow group velocity values of $v_g = c/7$ have been reported before close to the ENZ wavelength [26]. However, in the alternative current example of an active ENZ plasmonic waveguide with gain coefficient $\delta = 0.017$, the group velocity becomes much smaller and reaches to $v_g = c/570$ (stopped light regime) at the ENZ EP degeneracy, where $c$ is the velocity of light in free space. We have also considered the case of complex frequency excitation in the proposed active plasmonic devices leading to purely real and very low wavenumbers, which can retain their slow group velocity even in the presence of losses [62,63]. A more detailed analysis about the complex frequency excitation and its influence on the system's group velocity can be found in [36]. We believe that such slow group velocities at the ENZ EP may inspire new applications in temporal soliton excitation and in the design of slow light devices.

## 5. Conclusions

In this work, an active ENZ plasmonic waveguide system has been designed by using a gain medium embedded inside its nanochannels. The proposed plasmonic system mimics the behavior of a non-Hermitian Hamiltonian microscale photonic system but at the nanoscale and for subwavelength dimensions. It was demonstrated that the introduction of gain can lead to more exotic effects compared to just plasmonic loss compensation, such as reflectionless ENZ response at a formed EP spectral degeneracy and super scattering behavior at a spectral singularity point. The conditions where the reflectionless loss-compensated ENZ and super scattering points occur have been theoretically analyzed by using a robust transmission-line analytical model and numerically validated with full-wave simulations. In addition, we have investigated several unique applications of the proposed active ENZ plasmonic system, including coherent directional absorption and strong optical bistability or all-optical switching due to enhanced nonlinear effects. In particular, when the active ENZ waveguide is illuminated by two counter-propagating plane waves, the output power was modulated from amplification and super scattering to directional absorption just by using different phase offsets between the two incident waves. As a final remark towards the practical implementation of the proposed configuration, we would like to stress that the entire plasmonic ENZ waveguide system may be embedded in an active material, which can be used as a substrate and superstrate, without affecting the results presented in this work. The interesting functionalities of the proposed active ENZ plasmonic waveguides provide a new path to achieve strong light-matter interactions and spectral degeneracies in nanoscale dimensions.

**Acknowledgments**

This work was partially supported by the National Science Foundation (DMR-1709612) and the Nebraska Materials Research Science and Engineering Center (MRSEC) (grant No. DMR-1420645).

**Figures**

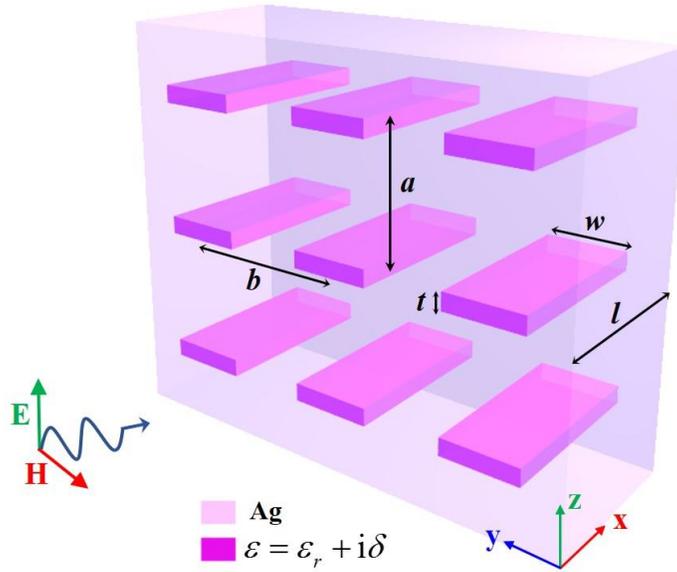

Figure 1 – Geometry of the periodic silver plasmonic waveguides. The rectangular slits are loaded with an active dielectric materials ($\varepsilon = \varepsilon_r + i\delta$) and are carved in the silver screen. The device is excited by a plane wave impinging at normal incidence.

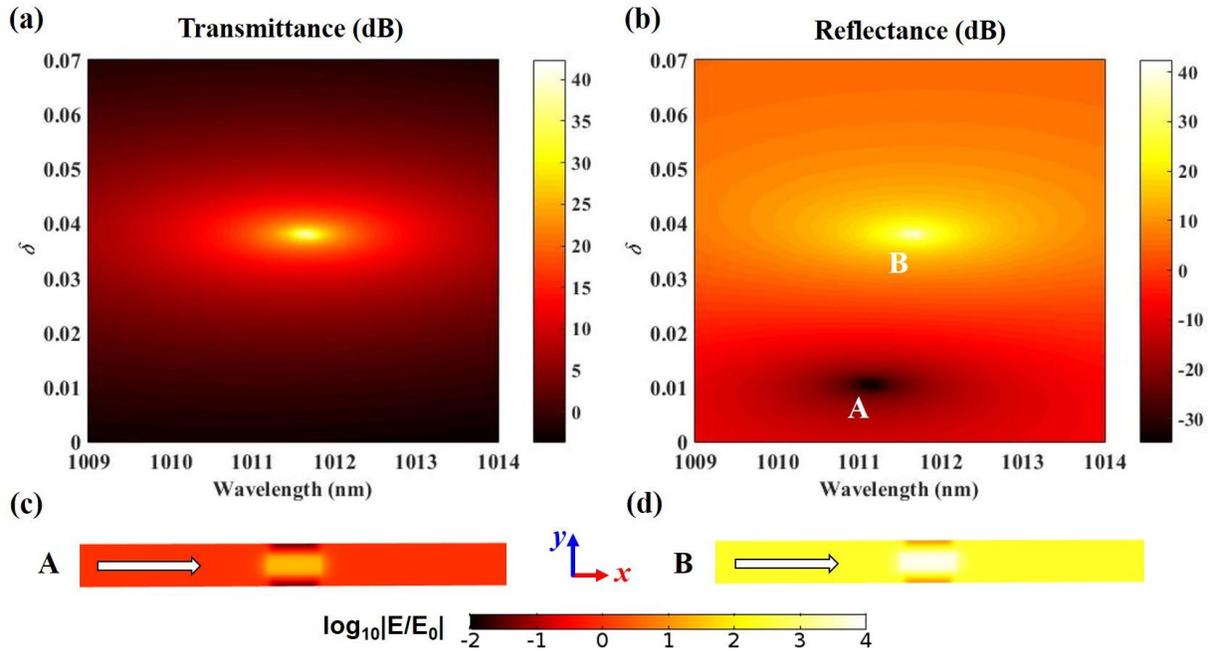

Figure 2 – (a) Transmittance and (b) reflectance as a function of the incident wavelength and the small imaginary part $\delta$ (gain) of the active dielectric material permittivity loaded in the waveguides. Reflectionless loss-compensated ENZ behavior is obtained at the exceptional point A. The super scattering (lasing) ENZ mode is obtained at the spectral singularity point B. Normalized electric field enhancement distribution in the channel's *xy*-plane operating at (c) the reflectionless loss-compensated ENZ point A and (d) the super scattering (lasing) ENZ point B. The white arrows depict the incident wave direction.

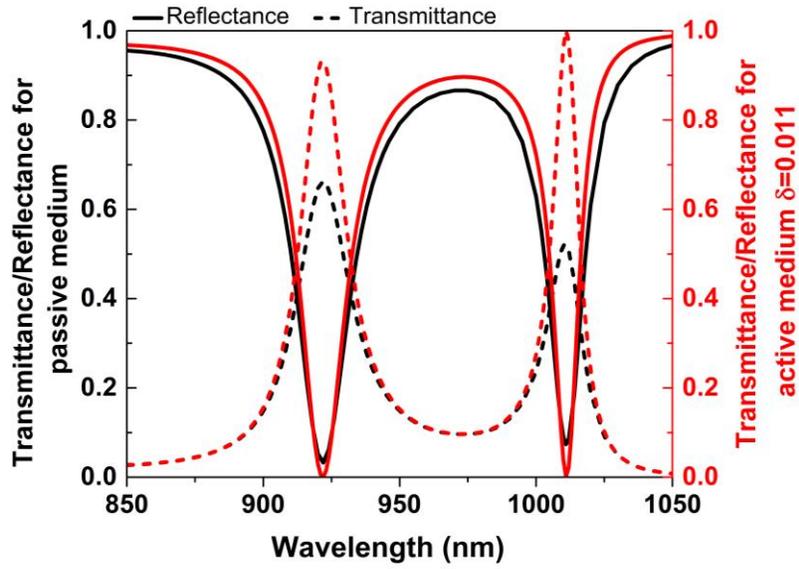

Figure 3 – Computed transmittance and reflectance of the plasmonic waveguide as a function of the incident wavelength with (red) and without (black) gain. An active dielectric material permittivity with imaginary part $\delta=0.011$ is used to obtain the reflectionless loss-compensated ENZ response shown by the exceptional point A in Fig. 2(b).

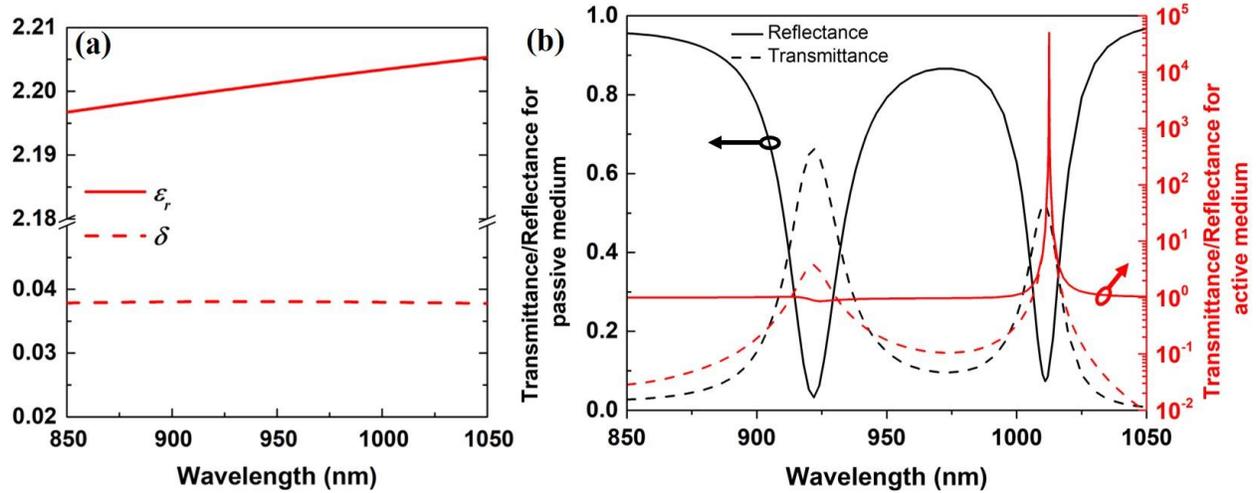

Figure 4 – (a) The Lorentz permittivity model of the dispersive realistic active (gain) dielectric material. (b) Transmittance and reflectance of the plasmonic waveguides as a function of the incident wavelength with (red) and without (black) gain. The permittivity of the active material in this case follows the Lorentz model shown in (a).

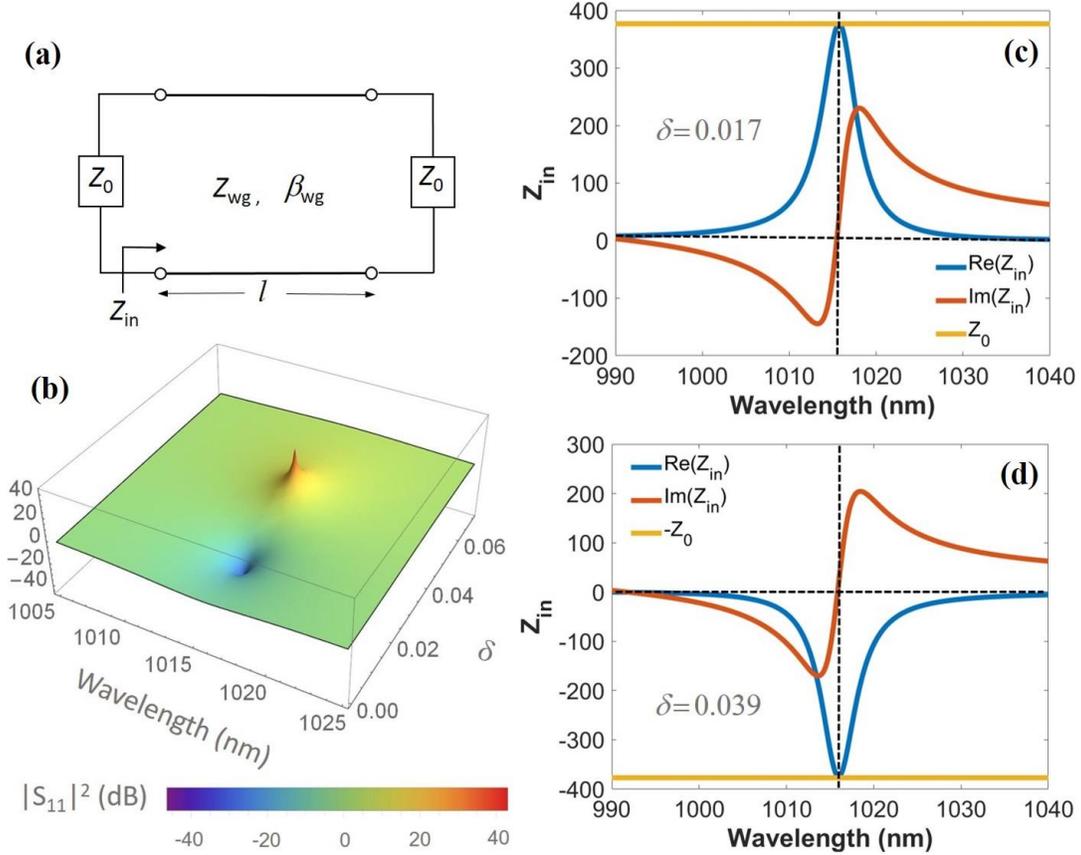

Figure 5 – (a) Transmission-line model of an active plasmonic waveguide $Z_{wg}$ terminated by two loads with impedances $Z_0$ due to the surrounding free space. (b) Analytically computed 3D distribution of the amplitude of the scattering parameter $|S_{11}|^2$ (reflectance) versus the incident wavelength and gain coefficient $\delta$. (c) The real and imaginary part of the input impedance $Z_{in}$ versus the incident wavelength for a fixed gain coefficient $\delta=0.017$. The condition of the reflectionless loss-compensated ENZ response given by Eq. (7) is perfectly satisfied at the exceptional point (vertical black dashed line). (d) The real and imaginary part of the input impedance $Z_{in}$ versus the incident wavelength for a fixed gain coefficient equal to $\delta=0.039$. The condition of the super scattering ENZ response given by Eq. (8) is perfectly satisfied at the spectral singularity point (vertical black dashed line).

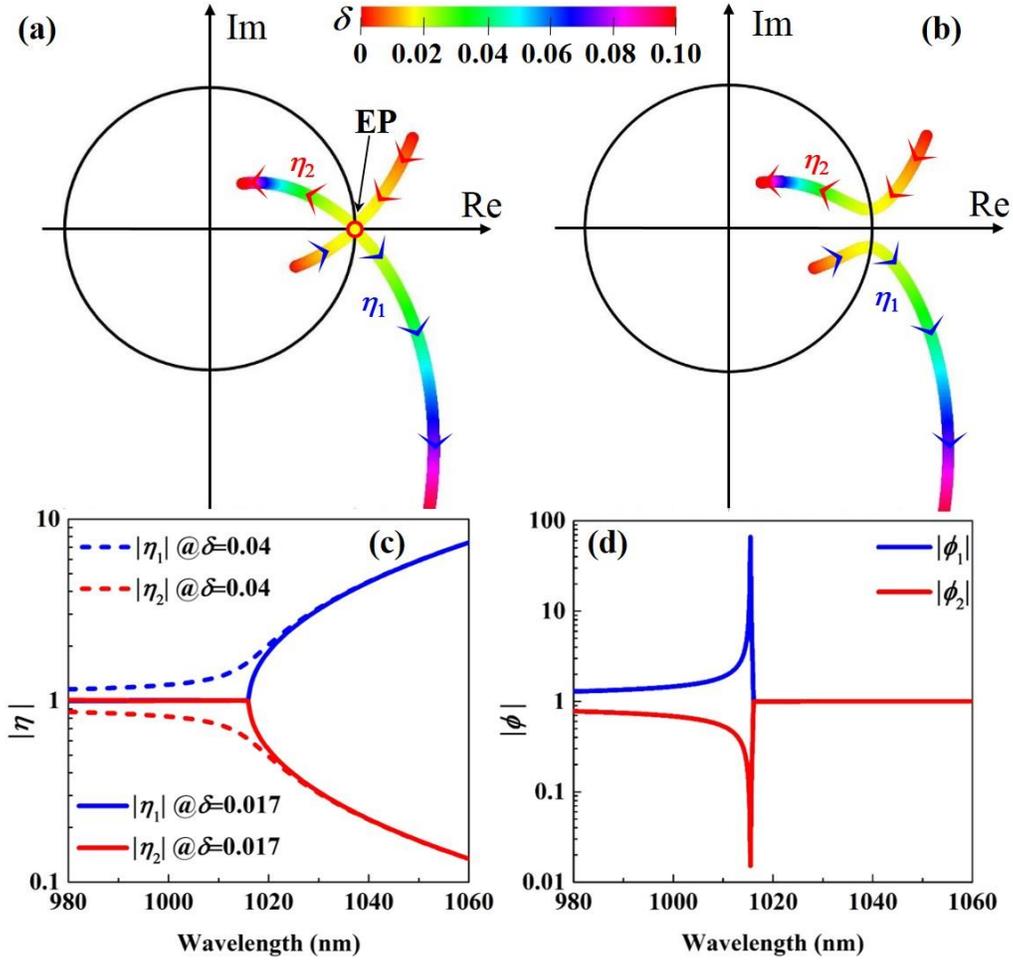

Figure 6 – (a), (b) Evolution of the two complex eigenvalues $\eta_{1,2}$ computed by Eq. (9) as a function of the gain coefficient $\delta$ at (a) the ENZ cut-off wavelength $(\lambda=1016\,\text{nm})$ and (b) slightly off the ENZ resonance wavelength $(\lambda=1015.8\,\text{nm})$. The two eigenvalues coalesce to form an EP degeneracy only at the ENZ cut-off wavelength which can be clearly seen in caption (a). (c) Absolute values of the two eigenvalues versus the wavelength for two different gain coefficients equal to $\delta$=0.017 (solid lines) and $\delta$=0.04 (dashed lines). (d) Absolute values of the two eigenvectors computed by Eq. (12) versus the wavelength for a fixed gain equal to $\delta$=0.017. A clear bifurcation point is observed for both eigenvalues [caption (c)] and eigenvectors [caption

(d)] for the gain value $\delta=0.017$. This bifurcation point coincides with the ENZ wavelength, which is a clear indication of an EP spectral degeneracy formation.

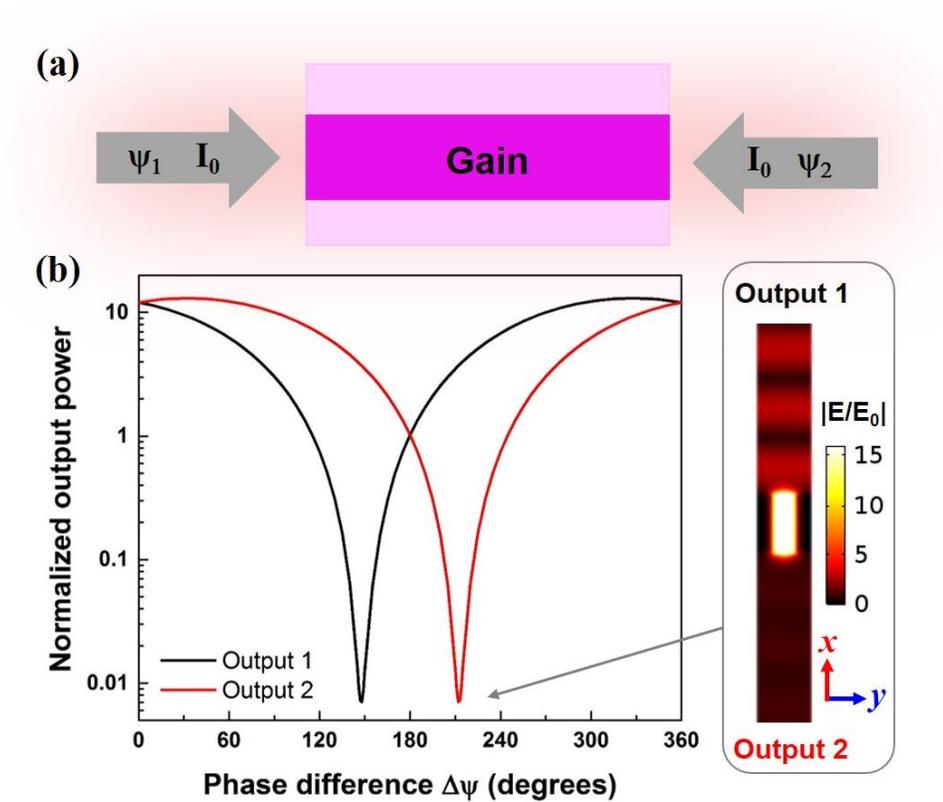

Figure 7 – (a) Active ENZ plasmonic waveguide illuminated by two counter-propagating plane waves. The phase difference of the two waves is $\Delta\psi=\psi_2-\psi_1$. (b) Normalized output power computed at both sides of the active plasmonic system versus the phase difference $\Delta\psi$ plotted slightly off the ENZ resonance wavelength ($\lambda=1009$ nm). The right inset caption demonstrates the amplitude of the normalized electric field distribution when $\Delta\psi=212°$. It can be clearly seen that the wave propagates only towards the output 1 in this case.

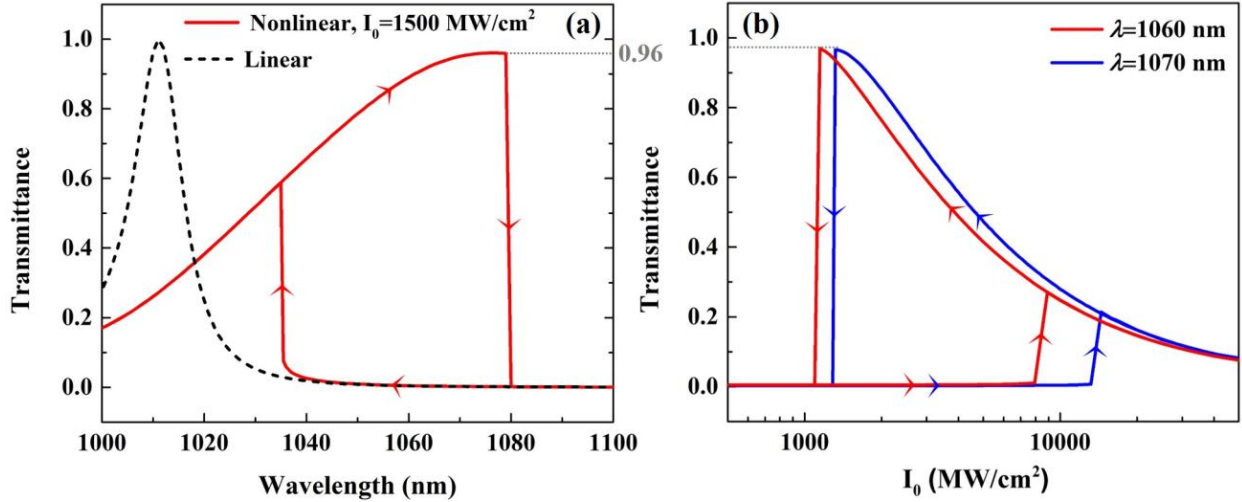

Figure 8 – Third-order optical nonlinear material introduced in the nanochannels of the active plasmonic waveguides. (a) Computed transmittance versus wavelength for both linear (dashed line) and nonlinear (solid line) operation when high input intensity $I_0 = 1500$ MW/cm$^2$ is applied. (b) Bistable transmission versus the input intensity $I_0$ computed for two different wavelengths both of them with slightly larger values compared to the ENZ EP wavelength.

# *Supplemental Material*: Exceptional points and spectral singularities in active epsilon-near-zero plasmonic waveguides

Ying Li and Christos Argyropoulos*

Dept. of Electrical & Computer Engineering, University of Nebraska-Lincoln, Lincoln, NE,

68588, USA

*christos.argyropoulos@unl.edu

## I. Effective epsilon-near-zero resonant response of plasmonic waveguides

In this work, a free-standing waveguide structure is proposed to obtain an effective epsilon-near-zero (ENZ) response. It is based on an array of plasmonic nanowaveguides operating close to their cut-off frequency. This anomalous quasi-static response was originally proposed in [1] and can lead to very strong fields inside the ENZ plasmonic nanowaveguides, characterized by uniform phase distribution and infinite phase velocity. It has been used for several applications in squeezing and tunneling light [1], enhancement of superradiance [2,3], boosting optical nonlinearities [4,5], exciting temporal solitons [6], and obtaining tunable coherent perfect absorbers [7].

In the following, we present a brief theoretical analysis of the proposed plasmonic nanowaveguides to prove and demonstrate that this plasmonic system can indeed realize an effective ENZ response at its cut-off frequency. As depicted in Fig. 1 in the main paper, periodic rectangular slits filled with dielectric material with width *w*, height $t \ll w$, length *l* and periods $a, b \gg t$, are carved in a sliver screen with relative Drude permittivity dispersion $\varepsilon_{Ag} = \varepsilon_\infty - f_p^2 / [f(f - i\gamma)]$, $f_p = 2175 \text{THz}$, $\gamma = 4.35 \text{THz}$, $\varepsilon_\infty = 5$ [8]. To make our analysis more simple, the dielectric slits are assumed to have no gain or loss (thus $\delta = 0$) and the relative permittivity of the dielectric material loaded in each slit is chosen to be $\varepsilon = \varepsilon_r = 2.2$. In addition, we always assume in the following an $\exp(i2\pi ft)$ time convention. Similar to the ENZ supercoupling effect presented in closed waveguides at optical frequencies [1] and microwave frequencies [9,10], the proposed approach to realize effective ENZ nanophotonic configurations consists of properly modulating the lateral width $w$ of the plasmonic channels (see Fig. 1 in main paper), such that each nanowaveguide operates at the cut-off of its dominant mode.

The lowest-order mode of such narrow orthogonal waveguide channels is a quasi-TE$_{10}$ mode, which is of a 'quasi' nature due to the non-negligible losses induced by the metallic waveguide walls at optical frequencies [4]. It can be evaluated by considering the transverse magnetic (TM) mode supported by the upper and lower plates of the nanochannel, in combination with the transverse electric (TE) mode confined between the lateral walls [11]. The TM mode supported from a parallel-plate waveguide formed by only the upper and lower metallic walls and a dielectric layer between them can be calculated by solving the following dispersion equation [12]:

$$\tanh\left(\sqrt{\beta_{pp}^2 - \varepsilon k_0^2}\,\frac{t}{2}\right) = -\frac{\varepsilon}{\varepsilon_{Ag}}\frac{\sqrt{\beta_{pp}^2 - k_{Ag}^2}}{\sqrt{\beta_{pp}^2 - \varepsilon k_0^2}}, \tag{S1}$$

where $k_{Ag} = k_0\sqrt{\varepsilon_{Ag}}$ is the wavenumber of silver, $k_0$ is the free space wavenumber, and $\beta_{pp}$ is the guided wavenumber of the parallel-plate waveguide formed by the upper and lower walls with the same dielectric layer height $t$, shown in Fig. 1 of the main paper, but infinite width $w$.

By considering the presence of lateral metallic walls spaced between a distance or width $w$, the TM mode is then transformed into a quasi-TE$_{10}$ mode inside the rectangular channel due to the modification in the guidance properties of the mode induced by the lateral metallic walls [13]. The dispersion equation of the rectangular waveguide configuration is given by [11]:

$$\tan\left(\sqrt{\beta_{pp}^2 - \beta_{wg}^2}\,\frac{w}{2}\right) = \frac{\sqrt{\beta_{wg}^2 - k_{Ag}^2}}{\sqrt{\beta_{pp}^2 - \beta_{wg}^2}},\tag{S2}$$

where $\beta_{wg}$ is the guided wavenumber of the dominant quasi-TE$_{10}$ mode of the proposed plasmonic waveguide system. Hence, the effective permittivity for the quasi-TE$_{10}$ mode of the entire array of plasmonic nanowaveguides system can be calculated by the formula [4]:

$$\varepsilon_{eff} = \frac{\beta_{wg}^2}{k_0^2} = \frac{\beta_{pp}^2}{k_0^2} - \frac{\pi^2 \varepsilon}{\left(\beta_{pp} w + 2\sqrt{\varepsilon}/\sqrt{\operatorname{Re}[-\varepsilon_{Ag}]}\right)^2}.\tag{S3}$$

In the limit $\varepsilon_{Ag} \to -\infty$ (assuming perfectly conducting metal) and when $\beta_{pp} = k_0\sqrt{\varepsilon}$, the effective permittivity can be simply written as:

$$\varepsilon_{eff} = \varepsilon - \frac{\pi^2}{k_0^2 w^2}\tag{S4}$$

As a result, the effective permittivity given by the approximate formula (S4) will become equal to zero leading to an effective ENZ resonant response only for a waveguide with width $w$ equal to: $w = \dfrac{\pi}{\sqrt{\varepsilon}\,k_0}$, which is the classic cut-off condition of rectangular waveguides at microwave frequencies. Therefore, the effective ENZ operation can be obtained at the cut-off wavelength $\left(\lambda_c = 2w\sqrt{\varepsilon}\right)$ of the dominant quasi-TE$_{10}$ mode, where $\operatorname{Re}[\varepsilon_{eff}] = \operatorname{Re}[\beta_{wg}] = 0$.

To further prove this point, we plot in Fig. S1(a) the normalized real part of the wavenumber versus the incident frequency for the proposed rectangular channel waveguide ($\beta_{wg}/k_0$, solid line) and an equivalent parallel plate waveguide ($\beta_{pp}/k_0$, dashed line) with similar dimensions. Both wavenumbers are normalized with the free space wavenumber $k_0$. It is clear that around 295 THz (corresponding to $\lambda = 1017 \text{nm}$), the proposed rectangular plasmonic waveguide system is at its cut-off condition, *i.e.* $\text{Re}[\beta_{wg}] = 0$, and its dominant quasi-TE$_{10}$ mode "feels" an effective near-zero permittivity [4]. Whereas in the case of the narrow parallel plate waveguide, $\beta_{pp}$ does not have a cut-off frequency $(\beta_{pp} \neq 0)$. Hence, this waveguide type cannot achieve ENZ operation at any frequency. In addition, we need to stress that the cut-off frequency is dependent only on the transverse cross-section of the proposed rectangular plasmonic waveguide and in particular on the nanoslit's width *w*, as it was shown in the previous paragraph. In Fig. S1(b), we record the variation of the cut-off frequency versus the slit width *w* for a fixed height *t* = 40nm of our waveguide system. It should be noted that by increasing the silt width *w*, the corresponding cut-off frequency decreases (redshift). This plot can be used to control and compute the cut-off frequency of the rectangular channels shown in Fig. 1, with the goal to efficiently tune and optimize the ENZ resonance frequency supported by the proposed plasmonic waveguide system.

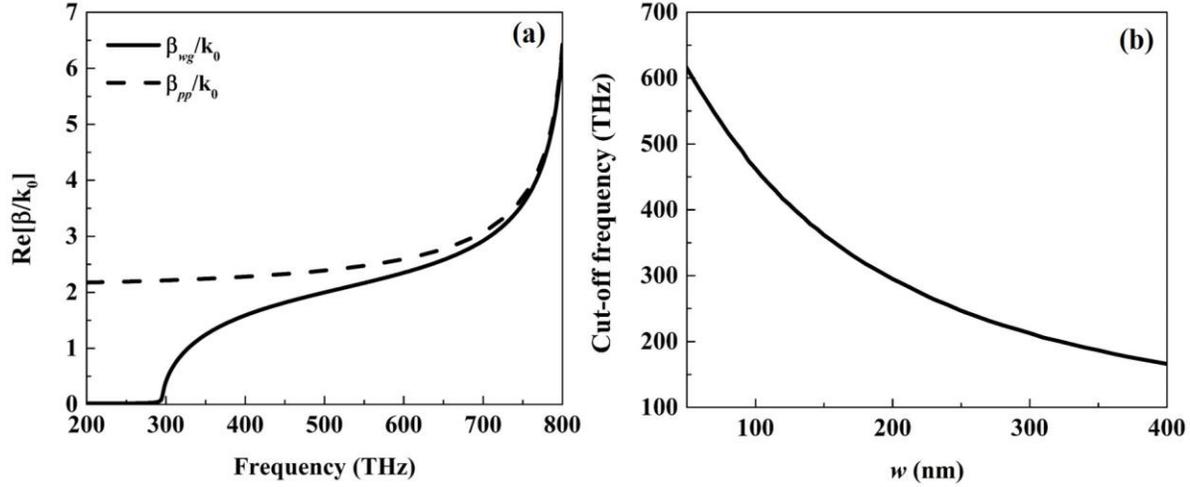

Figure S1 – (a) Dispersion of the real part of the guided wave number $\beta$ (normalized to the free-space wavevector $k_0$) versus frequency in a rectangular plasmonic channel with slit height $t = 40$nm and width $w = 200$nm (solid line), compared with the dispersion of an equivalent parallel-plate waveguide with the same height $t$ but infinite width $w$ (dashed line). (b) The computed cut-off frequency versus the slit width $w$ for fixed height $t = 40$nm of the proposed ENZ plasmonic waveguide.

In order to further clarify the ENZ resonance features in the proposed plasmonic waveguide system, the transmittance from the system has been evaluated by using full-wave numerical simulations (COMSOL Multiphysics, more details about simulations in the next section II). The waveguide parameters are chosen to obtain the cut-off frequency around $f = 295$THz (corresponding to $\lambda = 1017$nm), by using a slit height $t = 40$nm and width $w = 200$nm, consistent to the geometry used to produce the results in Fig. S1(a). The grating period is selected to be equal to $a = b = 400$nm. The transmittance results are computed and shown in Fig. S2 as a function of the incident wavelength for three different waveguide lengths: $l = 500$nm (black line), $l = 700$nm (green line), and $l = 1\mu$m (red line). A strong transmission peak around the wavelength $\lambda = 1012$nm is always present, independent to the choice of $l$, which further proves that this is the cut-off wavelength of this waveguide. This cut-off wavelength is very close to the theoretical value ($\lambda = 1017$nm) computed by the theoretical analysis in the previous paragraph.

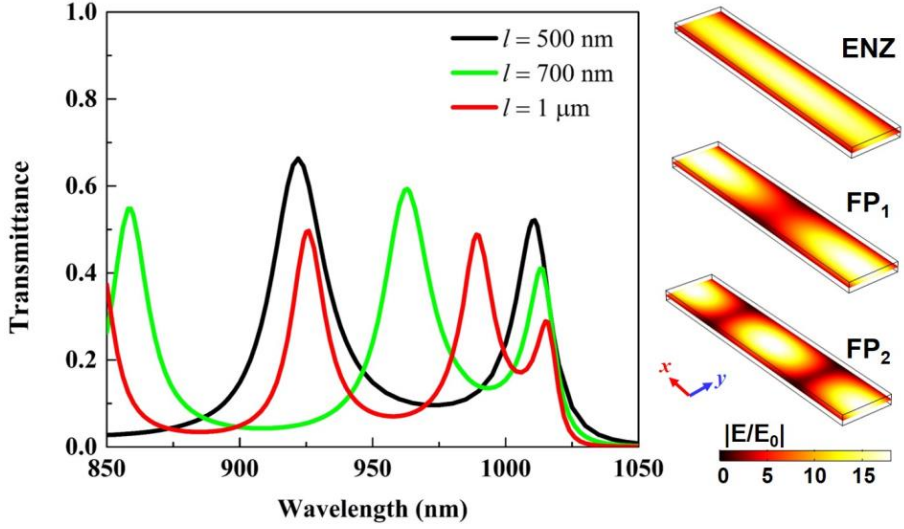

Figure S2 – Computed transmittance of the waveguide channels shown in Fig. 1 in the main paper with slit width $w = 200\,\text{nm}$, height $t = 40\,\text{nm}$, and grating period $a = b = 400\,\text{nm}$ for three different values of channel thicknesses: $l = 500\,\text{nm}$, $l = 700\,\text{nm}$, and $l = 1\,\mu\text{m}$. The right inset caption demonstrates the total electric field enhancement distribution in the channel's *xy*-plane operating at the ENZ, $FP_1$ and $FP_2$ resonant wavelengths. The channel thickness is fixed to $l = 1\,\mu\text{m}$ to compute these field distributions.

For longer wavelengths, the slits operate below cut-off and the incident waves are totally reflected back leading to zero transmittance, whereas for shorter wavelengths, several additional transmission peaks are observed that correspond to Fabry-Pérot (FP) resonances, which are strongly dependent on the waveguide's length *l*, as it can be seen in Fig. S2. We also plot the field enhancement distributions at the ENZ operation and other higher-order FP resonances for the case of $l = 1\mu\text{m}$ in the right insets of Fig. S2. It is interesting that at the ENZ cut-off wavelength, large and uniform field enhancement is obtained inside the nanochannels due to the effectively infinitely elongated guided wavelength in each nanowaveguide (tunneling effect), since $\lambda_g = 2\pi/\text{Re}\left[\beta_{wg}\right]$ and $\text{Re}\left[\beta_{wg}\right] = 0$ at ENZ. Whereas for higher-order FP resonances, the fields have the typical characteristics of standing wave distributions, where sharp minima and maxima are observed along the waveguide length. The enhanced and homogeneous fields at the ENZ resonance are responsible for the low required gain in order to excite and obtain the exceptional point and spectral singularity presented in the main paper.

## II. Numerical method details

During all our numerical computations, we employed the numerical simulation software COMSOL Multiphysics to solve the linear and nonlinear Maxwell's equations and investigate the unidirectional absorption and enhancement of third-order optical nonlinear effects by the proposed active ENZ plasmonic waveguides. COMSOL Multiphysics is a cross-platform Finite Element Method (FEM) based electromagnetic simulation software that is widely used by a plethora of research groups around the world to produce accurate linear and nonlinear simulation results. In particular, the RF Module of COMSOL Multiphysics was used for the electromagnetic analysis of all the currently presented simulation results. The FEM is primarily utilized to discretize the partial differential equations (PDEs) derived from the differential form of Maxwell's equations. In the following, we provide several additional details about the simulation procedures that were used to produce the results in the main paper.

**Geometry:** The geometry of the proposed array of plasmonic waveguides is shown in Fig. 1 in the main paper. It is composed of an array of narrow periodic rectangular slits carved in a sliver screen. Figure S3(a) depicts the numerical simulation domain of the proposed system that was used in our COMSOL simulations, which consists of a unit cell of the proposed plasmonic grating and air surrounding the structure. The plasmonic waveguide's length was chosen to be $l = 500\,\text{nm}$ in most results in the main paper and the length of the air domain was $l_{\text{Air}} = 750\,\text{nm}$. A cross-sectional view of the unit cell geometry is illustrated in Fig. S3(b). The slit dimensions are chosen to have width $w = 200\,\text{nm}$, height $t = 40\,\text{nm}$ (t≪w), and the grating period is selected to be equal to $a=b=400\text{nm}$. In order to simulate the proposed periodic grating geometry composed of an array of plasmonic waveguides, periodic boundary conditions (PBC) were employed in each side of the unit cell, along both *y*- and *z*-direction. The proposed plasmonic grating is illuminated by a normal incident *z*-polarized plane wave shown in Fig. S3(a), and port boundaries are placed at the front and back surfaces along the *x*-direction to create the incident plane wave. The PBC boundaries are indicated with yellow lines in Fig. S3(b) and the used ports are shown by dashed yellow lines in Fig. S3(a).

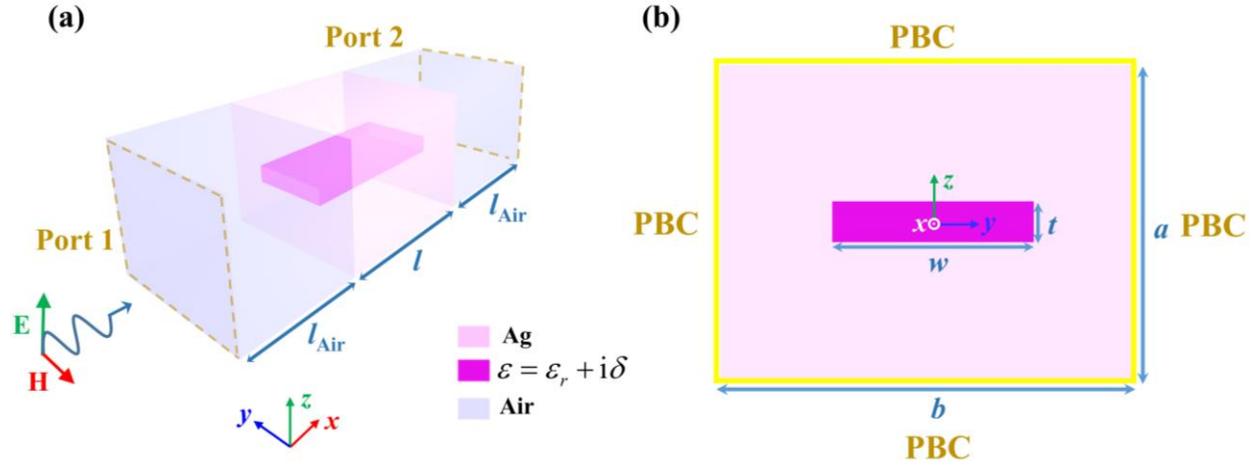

Figure S3 – (a) Schematic of the COMSOL simulation domain of the silver plasmonic nanowaveguide composed of a unit cell of the proposed plasmonic grating and air surrounding the structure. Port boundaries are placed at the back and front surfaces along the *x*-direction to generate a normal incident *z*-polarized plane wave. (b) Cross-sectional view of the unit cell geometry. PBCs are employed at each unit cell side.

**Materials:** We assume an $\exp(i2\pi ft)$ time convention during all our numerical simulations. The silver (Ag) permittivity dispersion values are taken from previously derived experimental data [8]. The slit in Fig. S3 is loaded with a dielectric material, whose relative permittivity value varies depending on the following different criteria:

a) Lossless material: The relative permittivity is equal to $\varepsilon = \varepsilon_r + i\delta$, where $\varepsilon_r = 2.2$ and $\delta = 0$. This means that the slit is loaded with a lossless dielectric material with relative permittivity $\varepsilon = \varepsilon_r = 2.2$.

b) Gain material: The relative permittivity is equal to $\varepsilon = \varepsilon_r + i\delta$, where the real part is $\varepsilon_r = 2.2$ and the imaginary part corresponds to the gain coefficient that is set to be equal to $\delta = 0.011$ in order to obtain the exceptional point (A) and $\delta = 0.038$ to observe the spectral singularity point (B) in Fig. 2 in the main paper. In addition, we also modeled the active dielectric medium using the more realistic and practical Lorentz dispersion active material model given by: $\varepsilon = \varepsilon_\infty + \varepsilon_{Lorentz}\omega_0^2/(\omega_0^2 - 2i\omega\delta_0 - \omega^2)$, where $\varepsilon_\infty = 2.175$, $\varepsilon_{Lorentz} = 0.06325$, $\omega_0 = 4.2\times10^{15}$ rad/s, $\delta_0 = 5.0\times10^{15}$ rad/s, and $\omega = 2\pi f$ (dispersion plotted in Fig. 4(a) in the main paper).

c) Kerr nonlinear material: The slit is loaded with a Kerr nonlinear dielectric material, which has a relative third-order nonlinear permittivity $\varepsilon_{ch} = \varepsilon + \chi^{(3)}|\mathbf{E}_{ch}|^2$. Here, $\varepsilon = 2.2 + 0.011i$ represents the linear permittivity of the active material with a gain coefficient $\delta = 0.011$ that is used to obtain the exceptional point (A) in Fig. 2 in the main paper. The third-order nonlinear susceptibility of silica is used in our nonlinear simulations that has relative low third-order nonlinear susceptibility equal to $\chi^{(3)} = 4.4\times10^{-20}\,\mathrm{m^2/V^2}$ [14]. Finally, $\mathbf{E}_{ch}$ is the magnitude of the local electric field induced inside the nanochannels.

**Results:** The reflectance and transmittance results, shown in Figs. 2-4 in the main paper, are computed as the ratio of the reflected and transmitted power divided by the incident power. These metrics can be directly computed by the S-parameters, which are built-in variables in the COMSOL RF Module when port boundaries are used. $S_{11}$ and $S_{22}$ are the reflection coefficients and $S_{21}$ and $S_{12}$ are the transmission coefficients from either port 1 or 2, respectively. The square of the amplitude of these coefficients gives the values of reflectance and transmittance that are computed and depicted in the main paper. Next, two counter-propagating plane waves with equal intensities $I_0$ were used to illuminate the proposed active plasmonic waveguides shown in the Fig. 7 scenario of the main paper, which is different from the one incident wave results presented in the previous Figs. 2-4. In this case, the normalized output power is computed as the summation of the output power normalized by the summation of the input power at both sides of the simulation domain (*i.e.* Port 1, Port 2), as shown in Fig. S3(a).

Finally, the bistable transmittance nonlinear responses shown in Fig. 8 in the main paper are computed by using nonlinear electromagnetic simulations always performed with COMSOL [7,15]. First, we compute the transmittance of the nonlinear ENZ active waveguides as the intensity of the input light increases. Each intensity step of our simulation is executed by using the solution calculated at the previous step. This initialization mechanism effectively enables us to retrieve the two stable branches of both hysteresis loops shown in Fig. 8 in the main paper. The first branch is obtained by starting our simulations with low-intensity values and gradually increasing the input intensity, while the second branch is computed by starting with high-intensity values and gradually decreasing the input intensity. Note that COMSOL nonlinear simulations are based on FEM and can efficiently compute the optical bistable response in a stable way, while taking into account the spatial variation of the local electric field and its impact on the intensity-dependent Kerr nonlinear permittivity. Consequently, the used COMSOL method is expected to be more accurate compared to other traditional semi-analytical and numerical nonlinear simulation methods used in the past, such as the graphical post-processing technique [15] and the finite-difference time-domain (FDTD) method [16]. More specifically, the graphical post-processing technique neglects how the electric field is modified by the optical nonlinear response and, instead, it assumes a spatially homogeneous dependence of the Kerr nonlinear permittivity on the local field. On the contrary, the FDTD method accounts for the inhomogeneous spatial distribution of the intensity-dependent Kerr nonlinear permittivity, similar to the currently used FEM-based COMSOL simulations. However, FDTD calculations typically require a heavy computational load in order to obtain accurate results. In addition, they are usually time-consuming, and, even more importantly, are prone to numerical instabilities, which is a major disadvantage compared to the stable nonlinear results obtained with the currently used FEM-based COMSOL simulations.

### III. Complex frequency excitation and group velocity calculations

We further study the case of complex frequency excitation and its influence on the real-wavevector mode. We also provide some additional details about the group velocity calculations. We first consider the proposed plasmonic nanowaveguides filled with dielectric material in their slits that is assumed to have no gain or loss (thus $\delta = 0$) and its relative permittivity is chosen to be $\varepsilon = \varepsilon_r = 2.2$. In this case, the imaginary part of the wavenumber Im($\beta$) as a function of the input complex frequency $(f = f_r + if_i)$ is shown in Fig. S4(a). Here, the real part of the excitation frequency is chosen to be fixed and equal to its cut-off frequency, *i.e.* $f_r = 295\,\text{THz}$, and the imaginary part is varied between the values of 0 to 5 THz. Note that in the current lossy ENZ plasmonic waveguide case ($\delta = 0$), the wavenumber is complex [$\text{Im}(\beta) \neq 0$] when using a real excitation frequency $(f_i = 0\,\text{point})$ due to the plasmonic losses stemming from the metallic waveguide walls. However, as we increase the imaginary part of the excitation frequency $f_i$, a point $(f_i = 1.1\,\text{THz})$ appears where the wavenumber becomes real [$\text{Im}(\beta) \simeq 0$] even in the presence of losses. The system's dispersion relations in the cases of real $(f_i = 0)$ and complex $(f_i = 1.1\,\text{THz})$ excitation frequencies are shown in Fig. S4(b). The waveguide system is at its cut-off condition at the ENZ resonance frequency but only in the case of the complex excitation frequency $\text{Re}[\beta] \simeq 0$ and $\text{Im}[\beta] \simeq 0$ can be achieved, because the passive ENZ system is lossy. Such purely real-wavevector modes, excited at the cut-off complex frequency $(f = 295 + 1.1i \text{ THz})$, can further decrease the group velocity to very low values around $v_g = c/240$.

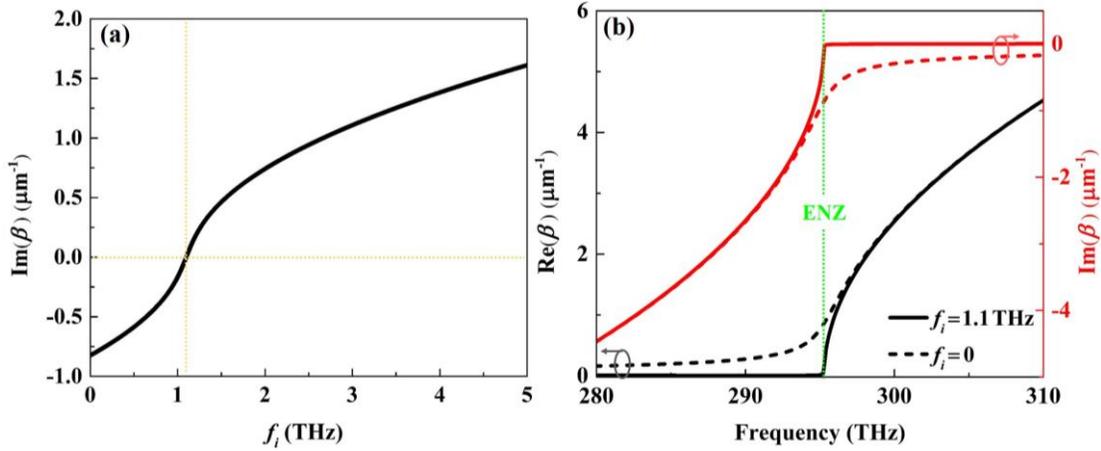

Figure S4 (a) Imaginary part of the computed wavenumber $\text{Im}(\beta)$ as a function of the imaginary part of the excitation frequency $f_i$. The real part of the excitation frequency is chosen to be equal to the ENZ cut-off frequency (295THz). In this case, the dielectric material loaded in the nanoslits has no gain or loss properties ($\delta$=0). (b). Real (black) and imaginary (red) parts of the wavenumber $\beta$ versus the real part of the excitation frequency with $(f_i = 1.1 \text{ THz})$ (solid line) or without $(f_i = 0 \text{ THz})$ (dashed line) imaginary part of the excitation frequency.

Next, we consider the proposed plasmonic waveguides to be filled with gain dielectric material (thus $\varepsilon = 2.2 + i\delta$). We plot in Fig. S5(a) the imaginary part of the wavenumber $\text{Im}(\beta)$ versus the gain coefficient $\delta$. The excitation frequency in this case is chosen to be real and is equal to the ENZ cut-off frequency $f = 295 \text{ THz}$. When $\delta = 0$ (no gain), the wavenumber is complex $[\text{Im}(\beta) \neq 0]$ due to the lossy nature of the proposed plasmonic ENZ nanostructures. When we increase the gain coefficient $\delta$, a special point appears (EP) for $\delta = 0.017$, where a purely real wavenumber is obtained by using a real excitation frequency. In this interesting point, which coincides with the presented in the paper EP, the ohmic losses induced by the metallic walls of the waveguides, in addition to the radiation losses of the proposed open non-Hermitian Hamiltonian system, are completely compensated by the active dielectric material loaded in the nanowaveguides. To further investigate the wavenumber behavior at the EP, the dispersion relations in the cases of passive ($\delta = 0$) and active ($\delta = 0.017$) dielectric slits are shown in Fig. S5(b). The waveguide system is at its cut-off condition at the ENZ resonance frequency but only in the active case both $\mathrm{Re}[\beta] \simeq 0$ and $\mathrm{Im}(\beta) \simeq 0$ can be achieved. Hence, the inherent loss of the plasmonic metallic waveguides and the active (gain) dielectric material loaded in them are perfectly balanced at this ENZ EP. By comparing the slope of the curves in Fig. S5(b), we find that the group velocity of the active plasmonic waveguide system becomes much smaller than the passive lossy system and can reach $v_g = c/570$ (stopped light regime) at the ENZ EP degeneracy.

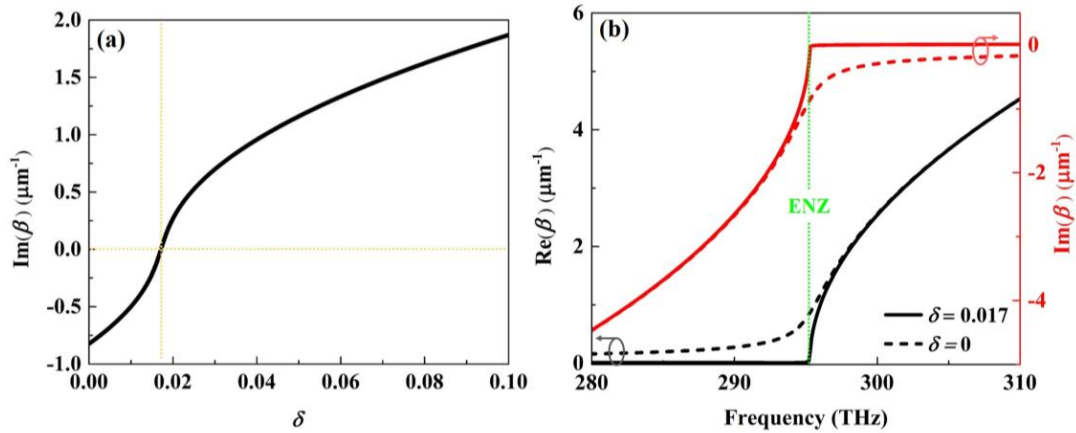

Figure S5. (a) Imaginary part of wavenumber Im($\beta$) as a function of gain coefficient $\delta$. The frequency is real and set to be equal to 295THz. (b) Real (black) and imaginary (red) parts of the wavenumber $\beta$ versus frequency with (solid) and without (dashed) gain.